\documentclass{eas}
\usepackage{graphicx}
\begin{document}

\title{The Cosmic Evolution of Gamma-Ray Burst Host Galaxies} 
\author{Sandra Savaglio}\address{MPI f. Extraterrestrial Physics, Garching bei M\"unchen, Germany}
\begin{abstract}

Due to their extreme luminosities, gamma-ray bursts (GRBs) can be detected in hostile regions of galaxies, nearby and at very high redshift, making them important cosmological probes. The investigation of galaxies hosting long-duration GRBs (whose progenitor is a massive star) demonstrated their connection to star formation. Still, the link to the total galaxy population is controversial, mainly because of the small-number statistics: $\sim1,100$ are the GRBs detected so far, $\sim280$ those with measured redshift, and $\sim70$ the hosts studied in detail. These are typically low-redshift ($z<1.5$), low luminosity, metal poor, and star-forming galaxies. On the other hand, at $1.5< z <4$, massive, metal rich and dusty, interacting galaxies are not uncommon. The most distant population ($z>4$) is poorly explored, but the deep limits reached point towards very small and star-forming objects, similar to the low-$z$ population. This `back to the future' behavior is a natural consequence of the connection of long GRBs to star formation in young regions of the universe.

\end{abstract}
\maketitle
\section{Introduction}

Long duration gamma-ray bursts (GRBs), the majority of known GRBs, are associated with the core collapse of massive stars ($M> 40$ M$_\odot$; Heger et al.\ 2003), preferentially located in regions experiencing immediate star formation (Fruchter et al.\ 2006).  They are so luminous in the $\gamma$-ray that they can shine through highly absorbed galaxies, normally difficult to see using conventional techniques. It is often claimed that GRB hosts are special galaxies, characterized by low chemical enrichment (e.g., Levesque et al.\ 2010). However, high metallicities have been measured in several hosts at $z>2$ (Savaglio 2012; and references therein) suggesting that intense star formation might be the dominant factor producing a GRB (Fynbo et al.\ 2008; Pontzen et al.\ 2010), rather than metallicity.

\begin{figure*}
\centering
\includegraphics[scale=0.5]{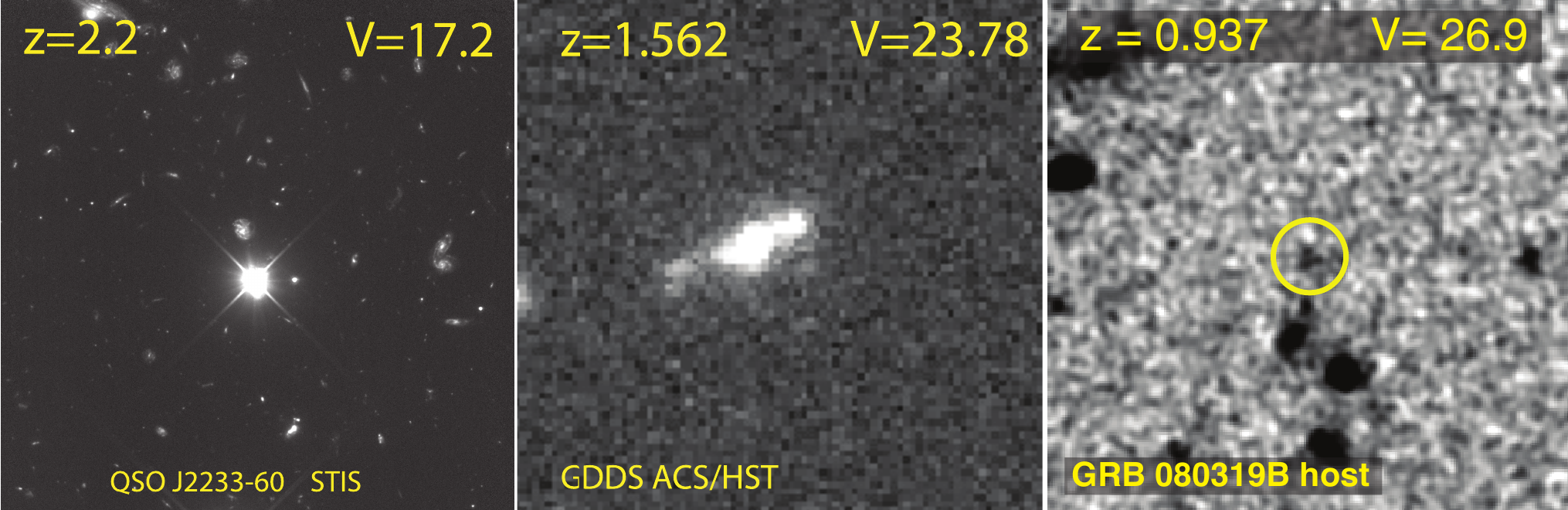}
\caption{\footnotesize{The ISM in the distant universe is investigated by observing different targets. {\it From left to right}: a bright ($m_V=17.2$) QSO, a faint ($m_V=23.8$) field galaxy and a very faint ($m_V=26.9$) GRB host galaxy. The GRB host is $\sim 8,000$ times fainter than the QSO.}}
\label{fqso}
\end{figure*}

In the past, the heavy element enrichment of the universe has been measured from the interstellar medium (ISM) of substantially different galaxy populations (Fig.\,\ref{fqso}). For a long time, absorption lines in the cold ISM (damped Lyman-$\alpha$ systems, DLAs) in bright QSO spectra were easily accessible to the highest redshift. More recently, 8-m class telescopes allowed the detection of emission lines in the hot ISM of star-forming galaxies up to $z\sim 3$ (Maiolino et al., 2008). Now, GRBs can probe both components of the ISM simultaneously, almost regardless of galaxy brightness. An extreme case is the host of GRB\,080319B ($z=0.937$, Fig.\,\ref{fqso}) characterized by a stellar mass $M_\ast \sim 5.5\times10^7$ M$_\odot$, metallicity $\log Z/Z_\odot \sim 0.7$, star formation rate SFR $=0.1$ M$_\odot$\,yr$^{-1}$, and specific star formation rate $s$SFR $=1.8$ Gyr$^{-1}$ (Tanvir et al.\ 2010). For comparison, the median values of a sample of 46 GRB hosts are: $z = 0.75$, $M_\ast = 2\times10^9$ M$_\odot$, SFR $= 2.5$ M$_\odot$\,yr$^{-1}$, and $s$SFR $= 1.25$ Gyr$^{-1}$ (Savaglio et al.\ 2009).

\section{The properties of GRB host galaxies}

\begin{figure}
\centering
\includegraphics[scale=0.43]{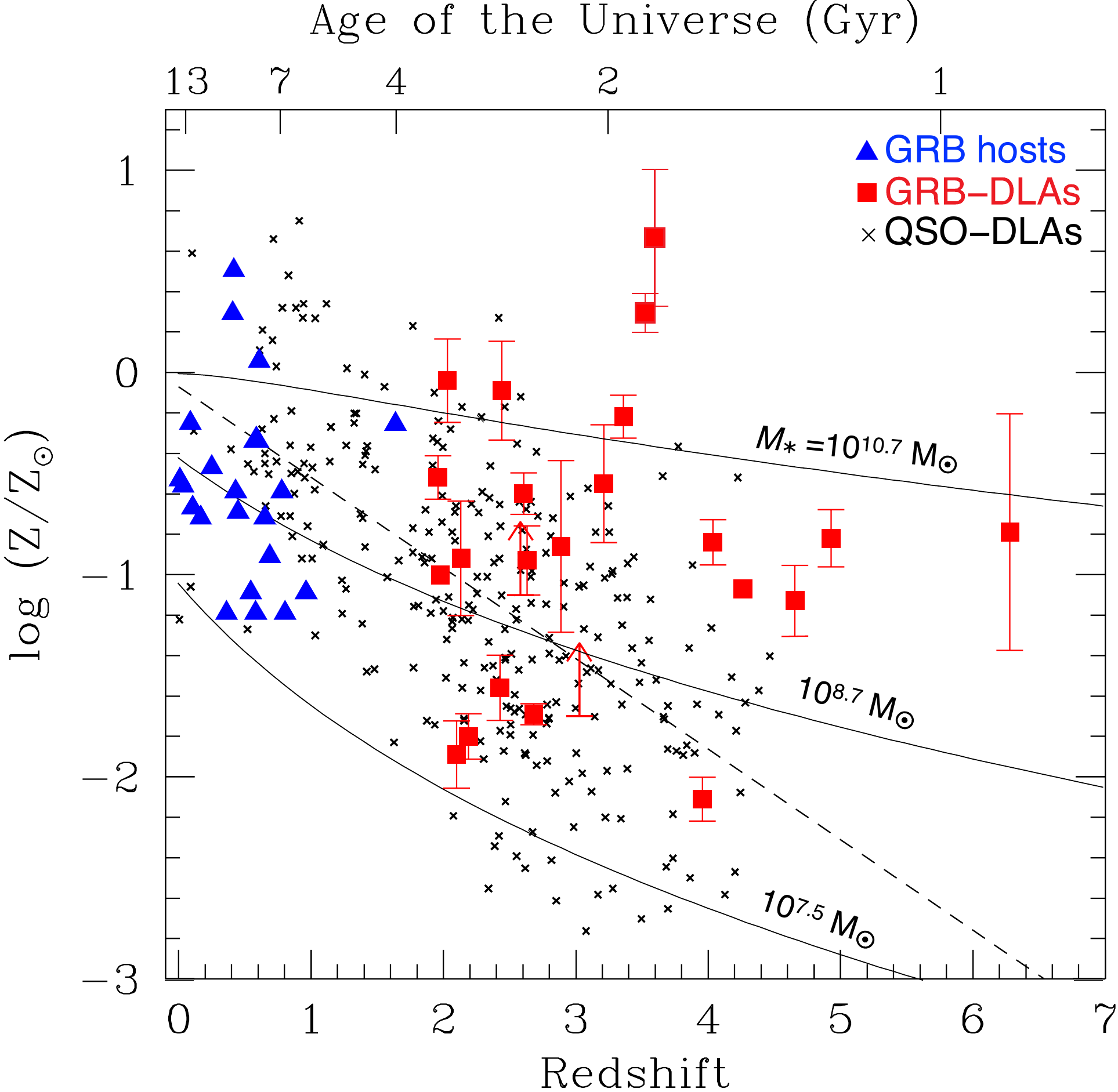}
\caption{\footnotesize{Metallicities of GRB-DLAs at $z>2$ (red squares; Savaglio et al.\ 2012; and references therein) and GRB hosts at $z<2$ (blue triangles; Savaglio et al.\ 2009; Levesque et al.\ 2010; Perley et al.\ 2012;  Kr{\"u}hler et al.\ 2012; Niino et al.\ 2012). With the exception of the metal-rich host of the short GRB\,100206 at $z=0.407$ (Perley et al.\ 2012), these are all long GRBs. QSO-DLAs are black crosses (dashed line: linear correlation). Solid curves are expected metallicities  of star-forming galaxies with different stellar masses, derived from the empirical mass-metallicity relation and its redshift evolution (Savaglio et al.\ 2005).}}
\label{Z}
\end{figure}

Not all GRB hosts are metal poor (Fig.\,\ref{Z}). At  $z <2$, a few  exceptions were found (Levesque et al.\ 2010; Perley et al.\ 2012;  Kr{\"u}hler et al.\ 2012; Niino et al.\ 2012). At $z>2$, GRB-DLAs display a large dispersion, between the $\sim1/100$ solar value in GRB\,090926A at $z=2.1062$ (Rau et al.\ 2010), and the gas-rich pair (separation $\sim 700$ km s$^{-1}$) in GRB\,090323 at $z=3.57$ (Savaglio et al.\ 2012). 

\begin{figure}
\centering
\includegraphics[scale=0.63]{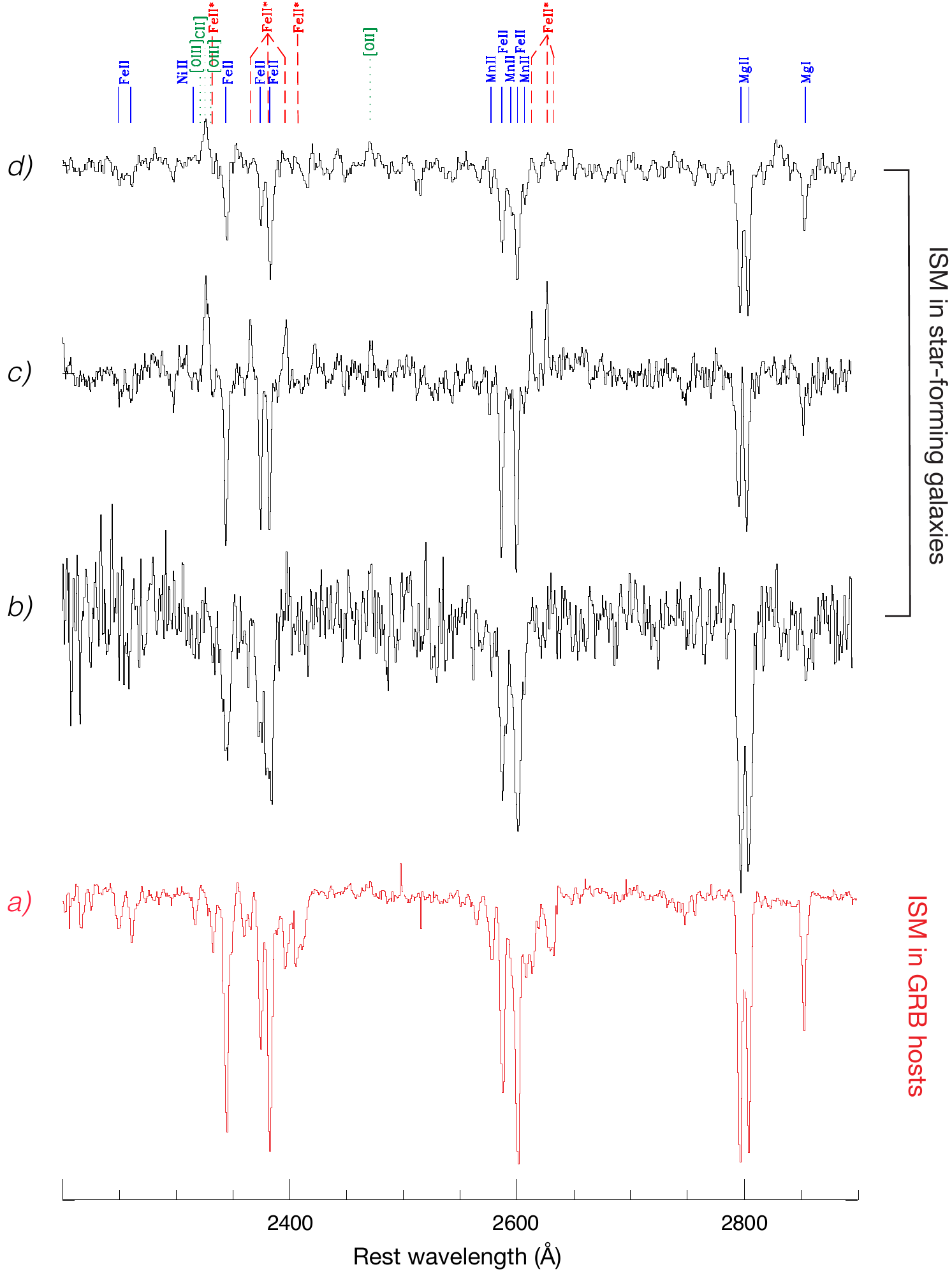}
\caption{\footnotesize{Composite UV spectra probing the ISM of galaxy populations, normalized and arbitrarily shifted on the $y$-axis for clarity. {\it From bottom to top}: $a)$ $z=0.9-1.5$ GRB afterglow composite (Christensen et al.\ 2011); $b)$ $z\sim1.6$ massive galaxies (Savaglio et al.\ 2004); $c)$ $z \sim 1$ UV bright galaxies (Martin et al. 2012); $d)$ local starburst and star-forming galaxies (Leitherer et al.\ 2011). Spectra are in scale  and shifted for clarity. Vertical marks indicate resonance (solid blue), fluorescent (dashed red), and nebular (dotted green) lines.}}
\label{composite}
\end{figure}

Fig.\,\ref{composite} displays an instructive composite of  $\sim25$ GRB-afterglow spectra in the interval $\lambda\lambda = 2200 - 2900$ \AA, mainly in the redshift interval $z=0.9-1.5$, at red and blue ends, respectively  (Christensen et al.\ 2011). The metallicity is not measured in the sample, but cold-ISM absorption lines are strong, especially in optically dark bursts (having an optical to X-ray spectral index $\beta_{\rm OX} < 0.5$). Moreover, the comparison in the same figure to the composite spectra of representative galaxy populations suggests that those GRB hosts cannot be metal poor and/or small. The average spectrum of 13 massive galaxies at $z \sim 1.6$ (median $M_\ast = 2.4\times10^{10}$ M$_\odot$, SFR $= 30$ M$_\odot$\,yr$^{-1}$, $s$SFR $= 1.2$ Gyr$^{-1}$; Savaglio et al.\ 2004) is very similar to the GRB composite. The one of a complete sample of UV-bright $z \sim1$ galaxies has much weaker  absorption lines, but sizable emission lines (from the hot gas), with the tendency of stronger absorbers to be more common in brighter galaxies (Martin et al.\ 2012). These galaxies have ${\rm SFR} =1-100$ M$_\odot$\,yr$^{-1}$ and $M_\ast = 10^{9.5}-10^{11.3}$ M$_\odot$\,yr$^{-1}$ ($s$SFR $ = 0.07 - 6$ Gyr$^{-1}$). A similar composite is the one of 28 local ($z<0.05$) starburst and star-forming galaxies, with median metallicity $\log Z/Z_\odot = -0.5$, UV luminosity and $K$-band absolute magnitude $L_{1500} = 5\times 10^{39}$ erg\,s$^{-1}$\,\AA$^{-1}$ and $M_K = -21.35$
($-25.1 < M_K < -15.4$; Leitherer et al.\ 2011). Using the empirical relations in Savaglio et al.\ (2009), we derive $M_\ast \sim 6\times10^{9}$ M$_\odot$, SFR$_{1500} \sim 1$ M$_\odot$\,yr$^{-1}$, and (assuming an optical extinction $A_V \sim 1$) 
an uncertain $s$SFR of a few Gyr$^{-1}$. Surprisingly enough, these values are not very dissimilar from those of the $z\sim0.75$ GRB host sample (Savaglio et al.\ 2009), despite the apparent difference with the GRB composite. We notice that the median redshift of the GRB host sample is lower than the redshift interval covered by the GRB composite ($z=0.9-1.5$), indicating again a redshift evolution of the galaxy population hosting GRBs.


\begin{figure}
\centering
\includegraphics[scale=0.42]{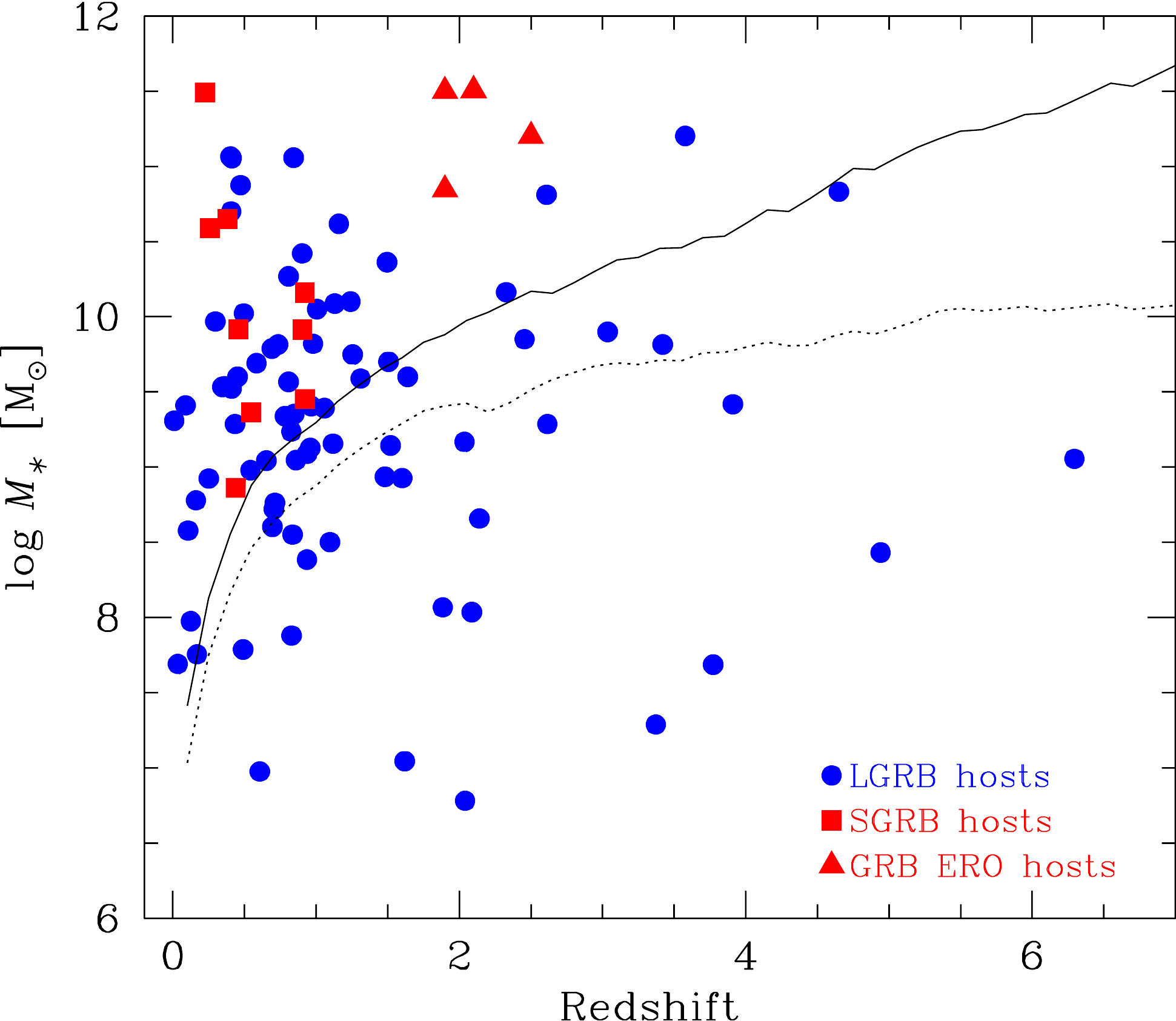}
\caption{\footnotesize{Stellar mass of GRB hosts as a function of redshift. Circles and squares are hosts associated with long and short GRBs, respectively (Savaglio et al., in prep.). Triangles are extremely red object (EROs) associated with optically dark GRBs (see Hunt et al.\ 2011 for details; Rossi et al.\ 2012). Solid and dashed lines represent the stellar mass of a galaxy with observed AB $K$-band magnitude $m_K = 24.3$, and old stellar population or constant SFR, respectively.}}
\label{mass2}
\end{figure}


\section{More massive GRB hosts at $z>1.5$}

At $z>1.5$, some GRB hosts are metal rich, massive (Fig.\,\ref{mass2}), dusty (dark GRBs), or highly star forming (Hunt et al.\ 2011; Kr\"uhler et al. 2011; Rossi et al.\ 2012). This kind of galaxies can be very bright in the sub-millimeter. However, latest study of a large sample at $z<1$ show not significant radio emission (Michalowski et al.\ 2012). One possible explanation is that the SFR in the low-$z$ universe does not occur mainly in sub-mm galaxies (SMGs), which can only account for at most 20\% of the cosmic SFR density (Michalowski et al.\ 2010). The steep redshift evolution indicates that future surveys can bridge the gap at $2 < z < 4$, and a sizable fraction of high-$z$ GRB hosts be SMGs. 

\begin{figure*}
\centering
\includegraphics[scale=0.6]{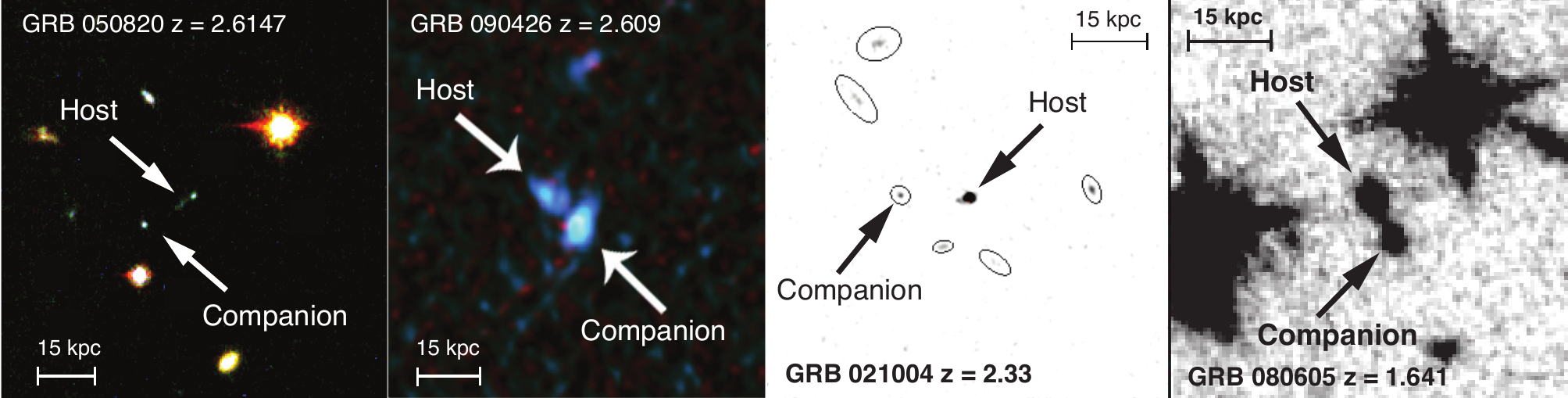}
\caption{\footnotesize{{\it From left to right}: the field of GRB\,050820 (Chen 2012), GRB\,090426 (Th{\"o}ne et al.\ 2011), GRB\,021004 (Fynbo et al.\ 2005) and GRB\,080605 (Kr{\"u}hler et al.\ 2012). All have a companion galaxy at the same redshift and separation $\leq 15$ kpc.}}
\label{fchen}
\end{figure*}

The fraction of pair absorbers in $z>1.5$ GRB afterglow spectra has been found to be almost three times higher than in QSO-DLAs (which probe random galaxies), suggesting that galaxy interactions may play a role in the formation of massive stars at high redshift (Savaglio et al.\ 2012; and references therein). Another indication is the large fraction (at least 40\%) of known GRB hosts at $z > 1.5$ showing interaction, disturbed morphologies, or galaxy pairs (Chen 2012; Kr\"uhler et al.\ 2012; Th\"oene et al.\ 2011; Vergani et al.\  2011). A few examples with a typical separation $10-15$ kpc is shown in Fig.\,\ref{fchen}. The interaction hypothesis is not surprising if one considers the higher fraction of galaxy mergers seen in the past of the universe with respect to today (Bluck et al.\ 2012).

\section{Ultra luminous supernovae at high redshift and the link to GRBs}

Galaxy mergers trigger instantaneous episodes of star formation or bursts. Thus, they are favorable sites of GRBs, and also ultra luminous supernovae (ULSNe). The latter are so bright that one day (when Extremely Large Telescopes will be operational) they will be used to explore the ISM at $z>1.5$, in territories difficult to achieve today with QSOs or GRBs.  

\begin{figure}
\centering
\includegraphics[scale=0.65]{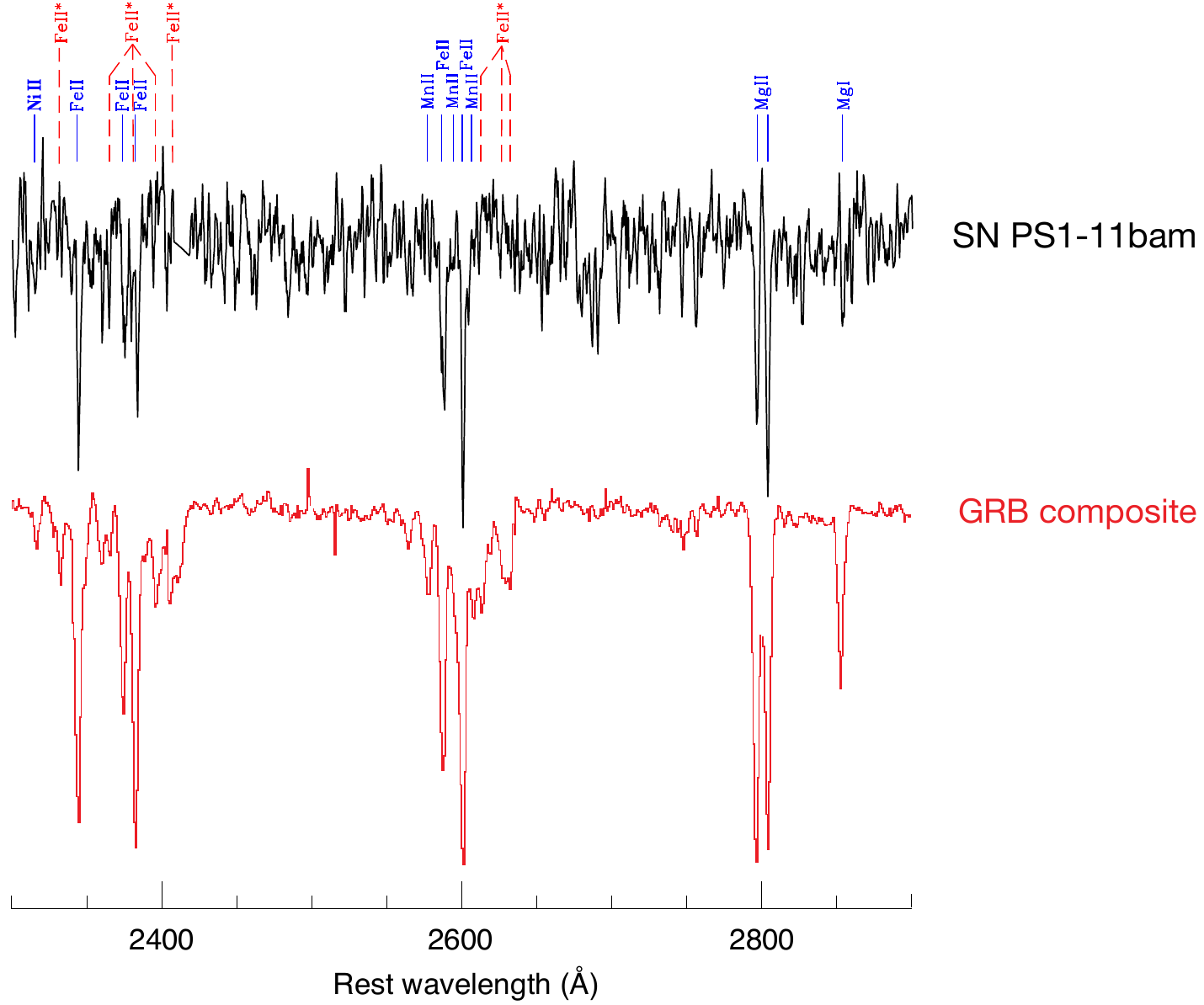}
\caption{\footnotesize{{\it Lower spectrum}: the GRB afterglow composite of Fig.\,\ref{composite}. {\it Upper spectrum}: ultra luminous supernova PS1-11bam at $z = 1.566$ (Berger et al.\ 2012). Its host has SFR $\sim10$ M$_\odot$\,yr$^{-1}$, stellar mass $M_\ast \sim 2\times10^9$ M$_\odot$ and $s$SFR $\sim 5$ Gry$^{-1}$.}}
\label{sn}
\end{figure}

There are more than one connection between GRBs and ULSNe. High-$z$ ULSNe, similar to type IIn and super-luminous supernovae (SLSNe, $M_g < -21.0$), are associated with very massive ($\sim 40-250$ M$_\odot$) and rare progenitors (Cooke et al.\ 2009; Gal-Yam 2012). ULSNe typically occur in faint, low mass and low metallicity galaxies (Neill et al. 2011) and have been detected from $z = 0.1$ to 1.6. The spectrum of one beautiful example, PS1-11bam at $z=1.566$, is in Fig.\,\ref{sn} (Berger et al.\ 2012). ULSNe are also unique in other ways. Unlike normal SN II or Ia, they are UV bright ($M_{\rm UV} \sim -20$ to $-23$) and, unlike GRBs, are bright for several months. They can be identified and followed up in the optical for $z > 1.5$ (Cooke et al.\ 2012), which has the advantage that they do not need a $\gamma$-ray or X-ray satellite, for fast identification and precise localization. GRBs can lose $3-4$ magnitudes in one day, and, at the moment, their discoveries are possible mainly thanks to the dedicated satellite {\it Swift}, which will be supported until 2014. 

Cooke et al.\ (2012) suggested that star forming episodes in interacting galaxies at $z>2$ increase the chance of observing the deaths of very a massive star. Moreover, Cooke et al.\ (2010) have shown that the number of close, interacting Lyman-break galaxy (LBG) pairs is higher than that expected from normal galaxy clustering, and that Ly$\alpha$ emission (a tracer of star formation) is anticorrelated with the separation of LBG pairs. These facts together suggest that LBG pairs are good sites of ULSNe, and hence GRBs.

\section{Back to the future: GRB hosts at $z>5$}

The investigation of GRB hosts at $z>4$ has been so far particularly difficult, which resulted only in rest-frame UV (e.g., star formation) detections. At $z>5$, only 5 GRB host fields have been observed (Basa et al.\ 2012; Tanvir et al.\ 2012), and no detection. If no dust correction is applied (dust is not expected to be  abundant in a $< 1$ Gyr-old universe), the UV-luminosity limit $L_{1500}$ can be translated into SFR$_{1500} <2.5$ M$_\odot$\,yr$^{-1}$ (Savaglio et al.\ 2009). Remarkable are the deep NIR observations with HST of the host of GRB\,090423 ($z=8.23$), which lead to $m_{\rm AB} > 30.29$ (Tanvir et al.\ 2012), or $L_{1500} < 3.7\times10^{38}$ erg s$^{-1}$ and SFR $ < 0.06$ M$_\odot$\,yr$^{-1}$.

\begin{figure}
\centering
\includegraphics[scale=0.38]{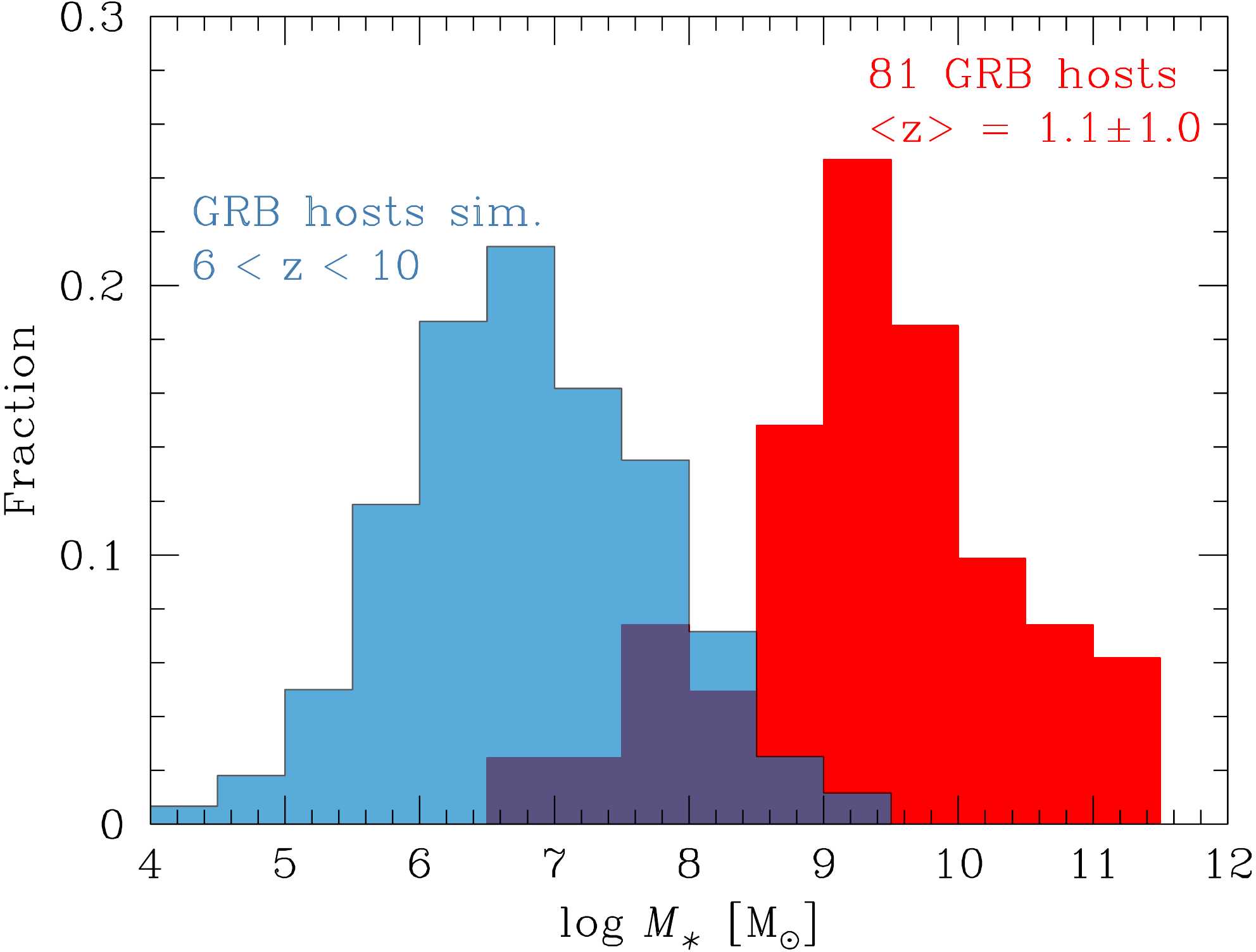}
\caption{\footnotesize{Fraction of GRB hosts with different stellar masses. {\it Left histogram}: $z > 6$ hosts of numerical simulations (Salvaterra et al.\ 2013); 70\% are the interval $M_\ast = 10^6-10^8$ M$_\odot$. {\it Right histogram}: 81 hosts at low redshift; 70\% are at $z<2$ (Savaglio et al., in prep.).}}
\label{mass}
\end{figure}

We can compare these low SFR limits to the stellar mass expected from numerical simulations. About 70\% of the hosts at $z>6$ predicted by Salvaterra et al.\ (2013) have stellar mass  in the range $M_\ast = 10^6-10^8$ M$_\odot$ (Fig.\,\ref{mass}), while star formation and metallicity are in the intervals SFR $=0.03-0.3$ M$_\odot$\,yr$^{-1}$ and $\log Z/Z_\odot = 0.01-0.1$. The comparison with a new observed sample at low redshift (81 GRB hosts, 70\% at $z<2$; Savaglio et al., in prep.) shows that in the past GRB hosts must have been really small. The SFR limit of the host of GRB\,090423 indicates that a very low stellar mass, $M_\ast \sim 10^6$ M$_\odot$, is possible if $s$SFR $<60$ Gyr$^{-1}$. Vice versa, if we assume $s$SFR $\sim 10$ Gyr$^{-1}$, the limit SFR $ < 0.06$ M$_\odot$\,yr$^{-1}$ would give $M_\ast < 6\times10^6$ M$_\odot$. 

In summary,  intrinsically faint GRB hosts are observed at $z <1.5$, whereas an important fraction are massive galaxies at intermediate redshift. Finally, in the primordial universe they are again likely small, star-forming, with no or little dust content.

\section{Conclusions}

The impact of GRB host galaxies on the understanding of galaxy formation and evolution is still affected by small number statistics. Their knowledge is mainly limited to the $z<1.5$ regime, where $\sim 60$ galaxies studied in detail point to a generally small, star forming and metal poor object. However, at $z>1.5$, metallicity, mass and dust extinction show a large spread, suggesting a different population. Additionally, about a half shows disturbed morphologies, interactions with nearby galaxies and mergers. All this is nicely connected to the idea that local massive ellipticals today were bursty in the past, with some of them experiencing close encounters with other galaxies, which likely triggered intense episodes of star formation. The recent discovery of ultra luminous supernovae at $z>1.5$, which have in common with GRBs a very massive progenitor, will certainly help understanding the nature of galaxies hosting these energetic events in the distant universe. At very high redshift, $z>5$, the situation might have changed again, because massive galaxies must have been very rare. At these distances, deep searches failed to detected any GRB host, and relatively low SFRs were inferred. Unless dust content was very high back then (unlikely), low SFRs means low galaxy mass. Therefore, GRB hosts in the past could have been more similar to the local counterparts.

\section*{Acknowledgments}

I thank Lise Christensen and Christy Tremonti for their help with composite spectra, and the workshop organizers for the kind invitation.


\end{document}